\documentclass[12pt]{article}
\usepackage{graphicx}
\usepackage{amsfonts}
\usepackage{amsmath}
\usepackage{color}

\title{Dynamic Network Centrality Summarizes Learning in the Human Brain}

\author{
Alexander V. Mantzaris\thanks{Department of Mathematics and Statistics, University of Strathclyde, UK}
\and
Danielle S. Bassett\thanks{Department of Physics, University of California, Santa Barbara, CA 93106, USA; Sage Center for the Study of the Mind, University of California, Santa Barbara, CA 93106, USA}
\and
Nicholas F. Wymbs\thanks{Department of Psychological \& Brain Sciences, University of California, Santa Barbara, CA 93106, USA}
\and
Ernesto Estrada\thanks{Department of Mathematics and Statistics, University of Strathclyde, UK}
\and
Mason A. Porter\thanks{Oxford Centre for Industrial and Applied Mathematics, Mathematical Institute,
University of Oxford, Oxford OX1 3LB, UK; CABDyN Complexity Centre, University of Oxford, Oxford, OX1 1HP, UK}
\and
Peter J. Mucha\thanks{Carolina Center for Interdisciplinary Applied Mathematics, Department of Mathematics, University of North Carolina, Chapel Hill, NC
27599-3250, USA; Institute for Advanced Materials, Nanoscience \& Technology, University of North Carolina, Chapel Hill, NC 27599-3216, USA}
\and
Scott T. Grafton \thanks{Department of Psychological \& Brain Sciences, University of California, Santa Barbara, CA 93106, USA}
\and
Desmond J. Higham\thanks{Department of Mathematics and Statistics, University of Strathclyde, UK}
}

\begin{document}

\maketitle

%\begin{flushleft}
%\textsf{
%\map{MAP: I ordered the authors by something that made sense to me.  Alex is obviously first.  I put other junior people (Dani and Nick) up.  I put Des last as "group leader" on this.  Aside from Alex being first, other junior people being in prominent places, and everybody else being somewhere (doesn't matter where) on the list, I will agree to any order---which in particular also includes putting me in any place in the order that you want.}
%\DB{I have added the rest of the authors and affiliations but have put them in no particular order.}
%\djh{DJH: I think we should keep most of the following text---in my experience reviewiers
%need to be told this explicitly, otherwise they assume it is regular paper, even if the editor tries to 
%flag it up.}

%\pre{Prepared as a \textbf{Report} for
%J.\ Roy.\ Soc.\ Interface: ``Reports are short,
%letter-style contributions (up to 2,500 words),
%which are published rapidly.
%Up to four displays
%(i.e. figures and tables) are allowed,
%of which no more than two
%should be figures.
%Preliminary and more speculative work of
%high-quality is strongly encouraged."}

%Current word count  including all \LaTeX\ commands
%is around 2,500, so we don't have a
%much extra space. (Can any coauthor accurately
%do a word count for
%this \LaTeX\ document?)
%}
%\end{flushleft}

%%%%%%%%%%%%%%

%\map{MAP: There was inconsistency in notation in the document.  I attempted to rectify this.  I don't think I caught everything.}

%%%%%%%%%%%

\newpage

\begin{abstract}

We study functional activity in the human brain using functional Magnetic Resonance Imaging and recently developed tools from network science. The data arise from the performance of a simple behavioural motor learning task. Unsupervised clustering of subjects with respect to similarity of network activity measured over three days of practice
produces significant evidence of `learning',  in the sense that subjects typically move between clusters (of subjects whose dynamics are similar) as time progresses. However, the high dimensionality and time-dependent nature of the data makes it difficult to explain which brain regions are driving this distinction. Using network centrality measures that respect the arrow of time, we express the data in an extremely compact form that characterizes the aggregate activity of each brain region in each experiment using a single coefficient, while reproducing information about learning that was discovered using the full data set. This compact summary allows key brain regions contributing to centrality to be visualized and interpreted. We thereby provide a proof of principle for the use of recently proposed dynamic centrality measures on temporal network data in neuroscience.

\end{abstract}

%%%%%%%%%%%%%%%%%%

\section{Motivation}\label{sec:mot}

A network-science perspective can give valuable insights into neuroscience data sets \cite{BuSp09}. In particular, it is useful for summarizing and comparing network properties in terms of a few key features \cite{CH09,DB09,RS10}, discovering cohesive groups of brain regions and other important patterns such as neuron orderings \cite{bassett11,CH10,HJB04}, and identifying important (i.e., `central') brain regions \cite{Sporns05,Z12}. 

Research on networks in neuroscience has focused primarily on static situations (because this allows one to use well-established tools \cite{newman10}), but current functional magnetic resonance imaging (fMRI) experiments offer the opportunity to study interactions that vary over time. Such \emph{dynamic} or \emph{temporal} networks arise in many other applications---including mobile phone communication \cite{jp07}, interactions in online social networks \cite{szell10}, criminal activities \cite{bertozzi}, voting in political bodies, and much more \cite{holme11}. 

One way to study temporal networks is to develop time-dependent generalizations of classical network `centrality' measures \cite{WF94,newman10}, which are designed to measure which nodes (or other network structures) are important in a network.  Different notions of centrality correspond to different contexts for what it means to be important.  The aim of the present paper is to test a recently proposed temporal centrality measure designed to quantify `communicability' in dynamic networks \cite{GHPE11} in the context of functional neuroscience and to highlight its potential for extracting useful information from this type of high-dimensional data.

The rest of this paper is organized as follows.  In Section \ref{sec:data}, we discuss the data that we use and what we aim to achieve in our analysis.  In Section \ref{sec:dyncom}, we review `communicability' in dynamic networks.  We present  our results in Section~\ref{sec:res}, and we discuss their implications in Section~\ref{sec:disc}.

%%%%%%%%%%%%%

\section{Data and Aims} \label{sec:data}

We study brain activity using the noninvasive neuroimaging technique of fMRI, which provides a quantitative measurement of regional changes in blood flow that are thought to be related to synaptic activity \cite{logothetis2001}. Our goal in this study is to identify meaningful temporal patterns related to brain function changing over a time scale of minutes to days. We therefore examine fMRI data that was acquired during a simple learning task in which $20$ subjects practiced short sequences of finger movements (12 movements per sequence type) over the course of 3 days \cite{bassett11,Wymbs2012}. Our data set is therefore composed of $60$ experimental sessions.

We construct dynamic functional brain networks by first parcellating the brain into $112$ anatomically distinct areas, which we represent as network nodes. We then partition the mean signal from each of these regions from all experimental sessions into $25$ time steps of roughly 3 minute duration each (corresponding to time series of 80 units in length). Thus, the full experiment consists of 25 time steps per subject. To estimate the interactions (i.e., the edge weights) between nodes, we need to calculate a measure of statistical similarity between regional activity profiles.  Using a wavelet transform, we extract frequency-specific activity from each time series in the range 0.06--0.12 Hz.  For each subject $s$, each experimental day $d$, each time step $t$, and each pair of regions $i$ and $j$, we define the weight of an edge connecting regions $i$ and $j$ as the coherence between the wavelet coefficient time series in each region (other measures of similarity are also possible \cite{smith2011}), and these weights form the elements of a weighted temporal network ${\bf A}$ with components $\left[{\bf  A}_{s,d}^{[t]}\right]_{i,j}$, where $s \in \{1,\cdots,20\}$, $d \in \{1,2,3\}$, and $t \in \{1,\cdots,25\}$.
%Here $ 1 \le s \le 20$, $ 1 \le d \le 3$ and $ 1 \le t \le 25$.  %MAP: these are discrete values, not integers, so we need to be precise about this.
We used a statistical correction (the false discovery rate) to threshold all connections for which we were not confident that the coherence value is significantly greater than that expected between random variables, and we then binarized the data. Specifically, $\left[{\bf A}_{s,d}^{[t]}\right]_{i,j} = 1$ if and only if the fMRI time series from regions $i$ and $j$ demonstrates statistically significant temporal coherence for subject $s$ on day $d$ at the $t^{\mathrm{th}}$ time step of the experiment.  We set all other $\left[{\bf A}_{s,d}^{[t]}\right]_{i,j}$ to $0$, from the measured connections (edges) approximately 9\% are kept as statistically significant. 
We can therefore view each experiment as a time-ordered sequence of $25$ binary, symmetric adjacency matrices of dimension $112 \times 112$.  We also note that each diagonal entry $\left[{\bf A}_{s,d}^{[t]}\right]_{i,i} = 0$.

In a previous examination of this data, we used time-dependent community detection \cite{Mucha2010} to identify statistically significant temporal evolution of network organization over time \cite{bassett11}. We found that network `flexibility', measured in terms of the time-varying allegiance of nodes to communities, in one experimental session predicted the relative amount of learning demonstrated in a future session. Our goal in the present work is to use a complementary approach, based on a recently proposed notion of temporal network centrality that respects the arrow of time \cite{GHPE11}, to examine the data from a different perspective.  Methods to study temporal networks are being developed rapidly, and they need testbed examples. It is therefore crucial to apply multiple viable approaches to the same data and evaluate the different insights and perspectives that they offer.  Very recent work has examined static centrality measures in functional brain networks \cite{kuhnert2012}, and our work generalizes such perspectives to time-dependent situations.

%%%%%%%%%%%%%

\section{Dynamic Communicability}\label{sec:dyncom}

Network centrality measures are designed to measure which nodes (or other network structures) are important \cite{WF94,newman10}, and many of them can be motivated by considering how information flows around a network \cite{Estradabook}. In such a perspective, central nodes are those that can use a network's connectivity structure to distribute or collect information effectively. In a time-varying network, where connections come and go, it is important to consider routes around the network that respect the arrow of time. For example, suppose that nodes $a$ and $b$ are connected today via an undirected edge and that nodes $b$ and $c$ are connected tomorrow via an undirected edge. The route $ a \mapsto b \mapsto c$ can thus be traversed over the course of the two days.  However, unless there are other edges, the reverse route from $c$ to $a$ cannot be taken, as the arrow of times introduces an asymmetry in the information flow \cite{holme11}. 

Reference \cite{GHPE11} quantified the ability of a node $i$ to send information to node $j$ across a time-dependent network by summing over all \emph{dynamic walks} from $i$ to $j$.
If we make the supposition that walks that traverse more edges are less relevant than those that traverse fewer edges, then the contribution to the sum from a walk that uses $w$ edges is scaled by $\alpha^w$. The parameter $\alpha \in (0,1)$ governs the extent to which we downweight for number of edges. This methodology was introduced by Katz
\cite{Katz53} for static, unweighted, undirected networks (also see the interpretation in \cite{bonacich87}). Katz noted that $\alpha$ can be interpreted as the independent probability that information successfully traverses an edge.  Importantly, because we binarize the data, we retain a dynamic analog of this interpretation in the networks that we study in the present paper.

The aforementioned pairwise summary, which suggests how effectively node $i$ can communicate with node $j$, is computed readily as the $(i,j)$ element in a product of matrix resolvents:
\begin{equation}\label{unnorm}
  {\cal P}_{s,d} :=
  \left( I - \alpha {\bf A}_{s,d}^{[1]} \right)^{-1}
   \left( I - \alpha {\bf A}_{s,d}^{[2]} \right)^{-1}
  \cdots
   \left( I - \alpha {\bf A}_{s,d}^{[25]} \right)^{-1}\,.
\end{equation}
In practice, we use a normalized version,
\begin{equation}
 {\cal Q}_{s,d} :=  \frac{
              {\cal P}_{s,d}
                   }
               {
  \| \,
          {\cal P}_{s,d}
   \, \|_2
 }\,,
\label{eq:Qdef}
\end{equation}
where $\| \cdot \|_2 $ denotes the Euclidean norm, in order
to avoid underflow and overflow.
The matrix inverses in (\ref{unnorm}) exist as long as $\alpha < \alpha^\star$, where $\alpha^\star$ is the reciprocal of the
maximum eigenvalue (in modulus) over all of the
individual adjacency matrices. In this work, we use
the value $\alpha = 0.9 \alpha^\star$.
%Computations in \cite{GHPE11}
%suggest that the results are not sensitive to the choice of
%$\alpha$, and 
As discussed in \cite{GHPE11}, averaging the connectivity
information over time and computing the Katz centrality
for this static summary (thereby ignoring the
time ordering) can produce significantly different
results.  Accordingly, it is important to use a method that respects the time-dependent nature of the problem.

The matrix entries $ \{ \left[ {\cal Q}_{s,d} \right]_{nj} \}_{j = 1}^{112}$ quantify the
ability of node $n$ to disseminate information to each node in a network.
One can sum over the elements in the $n^{\mathrm{th}}$ row to compute an
aggregate \emph{broadcast strength} ${\bf b}(n)$ for node $n$.
Similarly, by summing over the elements in the $n^{\mathrm{th}}$ column of $\left[ {\cal Q}_{s,d} \right]$, we quantify the ability of node $n$ to receive information using the \emph{receive strength} ${\bf r}(n)$.  This yields the broadcast centrality
\begin{align}
 {\bf b}(n)_{s,d} &:=
  \sum_{j = 1}^{112}
                        \left( {\cal Q}_{s,d} \right)_{nj}
\label{eq:br}                        
\end{align}
and receive centrality                        
\begin{align}                        
   {\bf r}(n)_{s,d} &:=
  \sum_{i = 1}^{112}
                      \left(  {\cal Q}_{s,d} \right)_{in}
\label{eq:re}
 \end{align}
from \cite{GHPE11}.

In the next section, we demonstrate that (a) the fMRI data provides evidence of learning, in the sense that subjects typically move between different clusters (of subjects based on similar neural activity patterns), based on these centrality values, as time progresses; and (b) that (time-respecting) dynamic communicability captures the same effect in a low-dimensional summary that is amenable to visualisation and interpretation.

%%%%%%%%%%%%%

\section{Results}\label{sec:res}

We seek brain signatures reflecting learning-related change as subjects repeat the motor task. We do this by treating each experimental session as a data point and performing unsupervised $k$-means clustering \cite{dataclustbook} to separate the experiments into two groups. We then view the clusters in terms of their subject/day identifications (IDs) to determine whether the two clusters
represent different stages of learning.  

In Table~\ref{tab:t1}(a), we show the
clustering that we obtain when we represent each experiment using its full set of
connectivity data---i.e., when we stack the matrices
 $\{ {\bf A}_{s,d}^{[t]} \}_{t=1}^{25}$ column by column into a single vector of dimension
$112 \times 112 \times 25 = 313,600$.
We hypothesize that cluster $2$ might represent a higher level of `ability' or `experience'
and hence that moving from cluster $1$ to cluster $2$ represents
the result of learning. To be more concrete, we regard the data processing/clustering
as a `success' for subject $s$ if the cluster label does
not decrease either between days $1$ and $2$ or between days $2$ and $3$.  In other words, a subject is successful if he/she does not exhibit a decrease in learning-related changes in brain function. For example, subject number $1$ in Table~\ref{tab:t1}(a) has the sequence $1,1,2$ and subject number $3$ has the sequence $1,2,2$, so both $s = 1$ and $s = 3$ are successful.  Subject number 9, who has the sequence $2,2,2$, is also successful.  However, subject 5 has the sequence $1,2,1$ and is therefore not successful.  In total, the $20$ subjects include $17$ successes and $3$ failures (subjects $5$, $8$, and $20$).  Using a permutation test, where we redistribute cluster labels $A$ and $B$ uniformly at random across experiments and map the labels $A$ and $B$ to the labels $1$ and $2$ in a way that minimizes the number of failures, we find that the achievement of $17$ or more successes has a p-value of $p \approx 0.0025$.  For this example (as well as all of our other $k$-means computations), we note that multiple runs with different starting values yield very similar results.

\begin{table}
{\scriptsize
\begin{center}
\begin{tabular}{ c |c | c | c | c | c | c | c | c | c | c | c | c | c | c | c | c | c | c | c | c | }
\multicolumn{21}{c}{\textbf{a) Full Temporal Data}}\\ \hline
  Subject ID       & 1 & 2 & 3 & 4 & 5 & 6 & 7 & 8 & 9 & 10 & 11 & 12 & 13 & 14 & 15 & 16 & 17 & 18 & 19 & 20 \\ \hline
Day 1 & 1 & 1 & 1 & 1 & 1 & 1 & 1 & 1 & 2 &  2  &  2 &   1  &   1 &   2 &   1 &   1 &   2 &  1  &   1 &  2  \\
Day 2 & 1 & 1 & 2 & 1 & 2 & 1 & 1 & 2 & 2 & 2   & 2  & 1    & 2   & 2   & 2   & 1   & 2   & 2   & 1   & 1 \\
Day 3 & 2 & 2 & 2 & 2 & 1 & 2 & 2 & 1 & 2 & 2   &  2 & 2    & 2   &  2  & 2   & 2   & 2   & 2   & 2   & 1
\end{tabular}
\\
\begin{tabular}{ c |c | c | c | c | c | c | c | c | c | c | c | c | c | c | c | c | c | c | c | c | }
\multicolumn{21}{c}{\textbf{b) Dynamic Communicability Matrix}}\\ \hline
    Subject ID      & 1 & 2 & 3 & 4 & 5 & 6 & 7 & 8 & 9 & 10 & 11 & 12 & 13 & 14 & 15 & 16 & 17 & 18 & 19 & 20 \\ \hline
Day 1 &   1 &    1   &  1 &    1    & 1  &   1    & 1 &    1    & 1  &   1   &  1  &   1  &   1  &   1    & 1 &    1   &  1&     1&     1&     1 \\
Day 2 &   1  &   1   &  2   &  1  &   2  &   1    & 1  &   1   &  2  &   1   &  2   &  1  &   1    & 1    & 2 &    1   &  2&     1&     1&     1 \\
Day 3 &   2 &    1    & 2   &  2  &   1    & 2 &    2  &   1 &    2  &   2  &   1    & 2&     2    & 1    & 2    & 2   &  1&     2&     2&     1
\end{tabular}
\end{center}
\caption{\label{tab:t1} Results of unsupervised $k$-means clustering into two groups, displayed by subject and by day, using (a) the full temporal data (which has dimension $313,600$) and (b) by summarizing each experiment in terms of the dynamic communicability matrix (\ref{eq:Qdef}) (which has dimension $112 \times 112 = 12,544$).  We also obtain the same results as in panel (b) using a vector of either the broadcast centralities (\ref{eq:br}) or receive centralities (\ref{eq:re}).  Each of these descriptions has a dimension of only $112$.
}}
\end{table}

In Table~\ref{tab:t1}(b), we show the corresponding results that we obtain when we summarize each experiment (of 25 time points) using its dynamic communicability matrix (\ref{eq:Qdef}), which we stack column by column into a vector of dimension $112 \times 112 = 12,544$.  We again observe $17$ successful subjects and $3$ failures (subjects $5$, $11$, and $17$).

We also apply the same clustering approach with each experiment collapsed to a vector of
either broadcast (\ref{eq:br}) or receive (\ref{eq:re}) centralities, where we recall that each component of either vector represents a single brain region.  Using either of these vectors, which have a dimension of only $112$, we find identical results as with the $12,544$-dimensional description.

The aforementioned results suggest (i) that there is evidence that the fMRI data
has captured a learning effect, and (ii) that the evidence remains intact even when we
vastly reduce the dimension of the data by using only broadcast or receive centrality measures,
which have  a natural interpretation in terms of quantifying the ability of a brain region to distribute or collect information.

\begin{figure}[t]
\begin{center}
\includegraphics[width=1\textwidth]{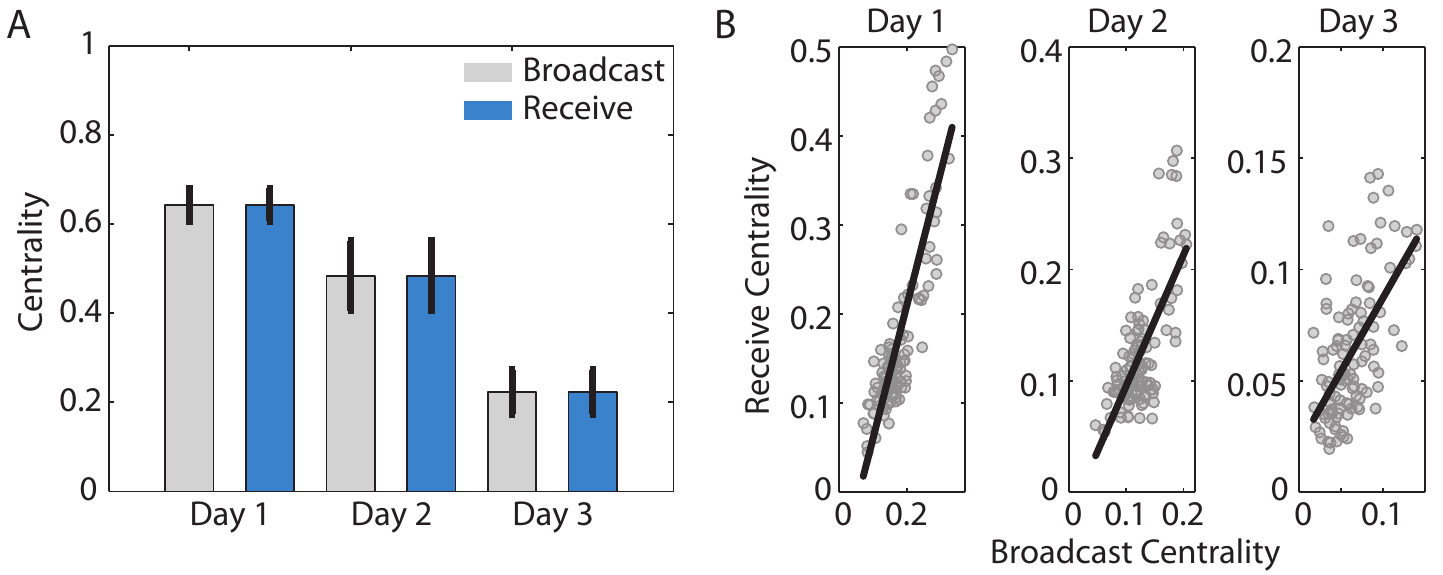}
\caption{Broadcast and Receive Centralities Change with Task Learning. \emph{(A)} Bar graph showing broadcast (light) and receive (dark) centralities averaged over brain regions for day 1, day 2, and day 3 of the experiment. Error bars indicate standard deviations of the mean over subjects. \emph{(B)} Scatterplots showing the Pearson correlations between broadcast and receive centralities for day 1 (correlation coefficient $r\approx0.86$; p-value $p \approx 2.4\times 10^{-35}$), day 2 ($r\approx0.71$; $p\approx6.6 \times 10^{-19}$), and day 3 ($r\approx0.60$; $p\approx1.8 \times 10^{-12}$) of the experiment.}
\label{fig:brain1}
\end{center}
\end{figure}

Because the broadcast and receive centralities relate directly to individual brain regions, we follow up on the results in Table~\ref{tab:t1} and study how these centralities vary over time. We find that broadcast and receive centralities both decrease appreciably over the 3 days of the experiment, suggesting their potential sensitivity to learning (see Fig.~\ref{fig:brain1}A). In addition to their temporal dependence, these two types of centrality vary over individuals in the experiment (see the error bars in Fig.~\ref{fig:brain1}A) and over brain regions (see Fig.~\ref{fig:brain1}A). We also find that broadcast and receive centralities are strongly correlated with one another over brain regions for all 3 days of the experiment (see Fig.~\ref{fig:brain1}B).

From the above results, it is unclear whether the broadcast and receive centralities for each brain region decrease similarly over days or whether the values for some brain regions decrease more than those for others. We therefore test whether any brain region has a change in its centrality values between day 1 and day 3 that is more than what is expected given the aggregate decrease shown in Fig.~\ref{fig:brain1}A. To do this, we normalize the broadcast and receive centrality vectors from the 60 communicability matrices separately. For each region, we then test whether the normalized centrality values differ significantly in day 1 versus day 3 using a permutation test in which we permuted the day 1 and day 3 labels uniformly at random. We find that no brain region demonstrates a significant decrease in either normalized broadcast centrality or normalized receive centrality from day 1 to day 3 ($p>0.01$; uncorrected for multiple comparisons).

\begin{figure}[t]
\begin{center}
\includegraphics[width=1\textwidth]{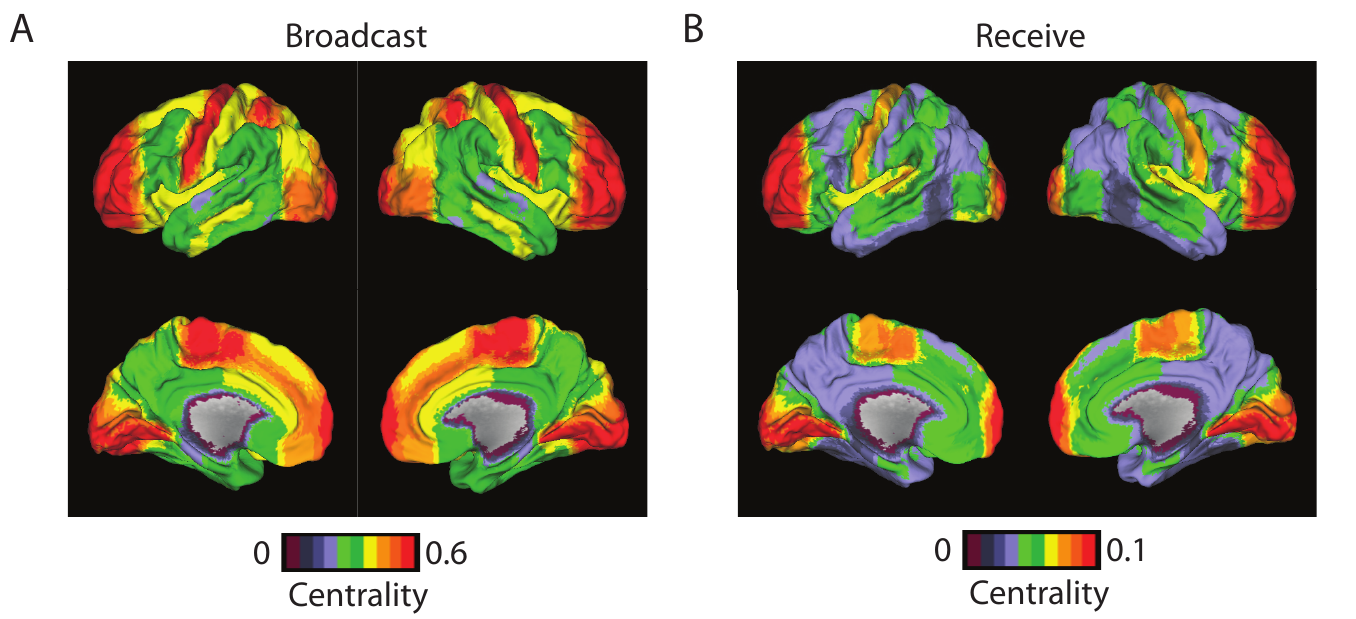}
\caption{Anatomical Distribution of Broadcast and Receive Centralities. Broadcast \emph{(A)} and receive \emph{(B)} centralities with normalized centrality vectors over subjects. }
\label{fig:brain2}
\end{center}
\end{figure}

In light of the result that brain regions do not differ significantly in the amount of centrality change with learning, one can study the anatomical distribution of centrality values by normalizing each experiment's centrality values, aggregating the normalized regional components from all the 60 experiments, and then viewing the aggregate over regions (see Fig.~\ref{fig:brain2}). We find that dynamic centrality values are greatest in bilateral precentral gyri (primary motor cortex), medial segment of the superior frontal gyrus (supplementary motor area), superior parietal lobule, and medial occipital cortices. 
This constellation of regions is a core sensorimotor system for controlling a broad range of visually guided actions \cite{Bernier2010}. Of particular note, two of these areas---the primary motor cortex and SMA---are consistently observed to demonstrate changes of local activity (\cite{Bischoff2004},\cite{Grafton1995},\cite{Hazeltine1997}) as well as changes of correlated activity during sequence learning \cite{Sun2007}.

%and \ref{fig:brain2}
% show results in the the case of
%broadcast and receive centrality, respectively,
%from (\ref{eq:br}).
%Here we show left and right hemispheres
%from the lateral and medial perspectives.
%To produce these pictures,
%for day $1$ we normalized the centrality
%measure across brain regions for each subject.
%In this way, each experiment has brain region
%centralities that sum to unity.
%For each brain region, we then sum over the $20$
%subjects.
%We repeated this for day $3$, and recorded the
%difference ``day three minus day one'' for each brain region.
%A larger value therefore indicates a greater increase
%in dynamic broadcast/receive centrality
%over time.

%%%%%%%%%%

\section{Discussion}
\label{sec:disc}

Broadcast and receive centralities are typically examined in the context of fast time scales, yet the 3-minute time scale in the experiments is a long time scale. The striking decrease in broadcast and receive centrality values over the course of the experiment is consistent with theoretical work indicating an increase in neural efficiency with learning \cite{Gobel2011}. 
In this view, greater skill at applying an initial task-related strategy leads to a temporal increase in the efficiency of neural processing, which can manifest as an aggregate decrease in measurements of brain function \cite{Landau2006}. 

It has been suggested that an understanding of such changes and their relationship to neural efficiency will require a more careful examination of functional connectivity patterns \cite{Kelly2005}, which are thought to be better indications of neuronal communication than activity patterns alone \cite{Sun2006}. The present paper highlights the potentially important effect of temporal dynamics in the consideration of neural efficiency, and recent methodological advances for the investigation of time-dependent networks \cite{holme11} now make it possible to pursue such efforts. We demonstrate decreases in functional connectivity patterns with task practice in early motor learning, suggesting that the brain might require less communication between distributed functional networks as skills become more automatic.

%\map{I kept 'practice' above because it seemed to make more sense than 'learning'; Scott, Nick: please confirm this is ok}

The ideas that we have employed in this study have important potential applications not only in the setting of fMRI experiments but also in the examination of functional neuroscience data using other experimental modes. Interest in studying whole-brain functional connectivity patterns in a network framework is growing  steadily \cite{BuSp09}, in part because network science includes a large set of diagnostics that are built to directly examine system connectivity and can be used to characterize the brain's structural organization. However, a study of the true dynamic nature of the brain requires the use of dynamic network diagnostics, the development of which is still in its early stages \cite{holme11}. We have demonstrated in this paper that broadcast and receive centralities are useful diagnostics for the study of temporal brain networks, and they have the additional advantage of respecting the arrow of time.  We expect such dynamic centrality measures to be similarly insightful in a wide variety of systems.

%\DB{\textbf{DB: May want to add more about the potential uses of these tools for other systems more broadly.}}

% \textsf{DISCUSS THE CARET PICTURES AND TALK ABOUT}

%\textsf{Do we need all these subfigures or
%can we get away with a subset?
%Should we add a table to accompany the
%pictures, e.g.
%listing the top 10 or 20 brain regions that
%have increased and decreased the most for
%broadcast and
%for receive, for each hemisphere?
%Maybe that would make it easier
%for the neuroscientists to add interpretation.
%}

\bigskip

\noindent
\textbf{Acknowledgements} This work was supported by the Errett Fisher Foundation,
the Templeton Foundation, David and Lucile Packard Foundation, PHS Grant
NS44393, Sage Center for the Study of the Mind, Institute for Collaborative
Biotechnologies through contract (\# W911NF-09-D-0001) from the U.S. Army
Research Office, and NSF (DMS-0645369). MAP acknowledges a research award (\#220020177) from the James S. McDonnell Foundation. 
AVM and DJH were supported by the 
Engineering and Physical Sciences Research Council/Research Councils UK Digital Economy programme through the MOLTEN (Mathematics of Large Technological Evolving Networks)
project EP/I016058/1.

 \bibliographystyle{siam}
\bibliography{mrefs-v6}

\end{document}